\documentclass[9pt,twocolumn,twoside]{osajnl}
\journal{ol} % Choose journal (ao, aop, josaa, josab, ol)

\setboolean{shortarticle}{true} 
\title{Spatially-resolved control of fictitious magnetic fields in a cold
atomic ensemble}

\author[1]{Adam Leszczy\'nski}
\author[1]{Mateusz Mazelanik}
\author[1]{Micha\l{} Lipka}
\author[1,*]{Micha\l{} Parniak}
\author[1]{Micha\l{} D\k{a}browski}
\author[1]{Wojciech Wasilewski}

\affil[1]{Institute of Experimental Physics, Faculty of Physics, University of Warsaw, Pasteura 5, 02-093 Warsaw, Poland}

\affil[*]{Corresponding author: michal.parniak@fuw.edu.pl}

\ociscodes{(020.0020) Atomic and molecular physics; (020.6580) Stark effect.}

\begin{abstract}
Effective and unrestricted engineering of atom-photon interactions requires precise spatially-resolved control of light beams. The significant potential of such manipulations lies in a set of disciplines ranging from solid state to atomic physics. Here we use a Zeeman-like ac-Stark shift of a shaped laser beam to perform rotations of spins with spatial resolution in a large ensemble of cold rubidium atoms. We show that inhomogeneities of light intensity are the main source of dephasing and thus decoherence, yet with proper beam shaping this deleterious effect is strongly mitigated allowing rotations of 15 rad within one spin-precession lifetime. Finally, as a particular example of a complex manipulation enabled by our scheme, we demonstrate a range of collapse-and-revival behaviours of a free-induction decay signal by imprinting comb-like patterns on the atomic ensemble.
\end{abstract}

\setboolean{displaycopyright}{true}

\begin{document}

\maketitle

The prospect of the all-optical arbitrary manipulation of spin drives
both classical and quantum engineering \cite{Yale2013,Moriyasu2009},
as it could enable efficient quantum information processing, e.g.
with single spins \cite{Goryca2014,Reiter2009}, as well as dense
and efficient storage of classical information \cite{Stupakiewicz2017}.
Generation of fictitious magnetic fields \cite{Cohen-Tannoudji1972}
by an optically-induced vector ac-Stark shift is a viable way of performing
such spin manipulations \cite{Moriyasu2009,Park2001}, due to its inherently
off-resonant and thus absorption-free nature. Moreover, application of fictitious
magnetic fields to spin ensembles proves to be a feasible way to reach
high sensitivities to real magnetic fields by means of all-optical
methods \cite{Zhivun2014,Lin2017,Sun2017}. Spatial control of the
applied effective potential paves the way towards novel applications
such as high spatial-resolution magnetometry \cite{Vengalattore2006},
magnetic field imaging \cite{Koschorreck2011}, magnetic gradiometry
\cite{Deb2013}, super-resolved imaging \cite{Hemmer2012} or implicitly generation
of tunable gauge potentials on ultracold atoms \cite{Goldman2014}.
Precise spatial control could also enable efficient operation of photon
echos \cite{Rosatzin1990,Zielonkowski1998} used in gradient-echo
quantum memories \cite{Sparkes2010,Chaneliere2015},
precise atom manipulation \cite{Park2001,Park2002} as well as
novel atom trapping techniques \cite{Schneeweiss2014,Albrecht2016}. New ways to
engineer efficient nonlinear light-atom interactions also arise as
the spatial degree of freedom inherently encompasses phase-matching
\cite{Parniak2015,Leszczynski2017}, also in quantum memories \cite{Zhang2014,Mazelanik2016,Parniak2016a}.

In this Letter we demonstrate spatially-resolved control of a vector
ac-Stark shift \cite{DeEchaniz2008} on a cold rubidium ensemble.
Using a phase-only spatial light modulator (SLM) we shape the spatial profile of
an off-resonant laser pulse that induces the ac-Stark shift on the
atoms. To characterize the interaction we observe the effect of the applied fictitious magnetic field
on a free-induction decay (FID) signal \cite{Smith2011,Behbood2013a}.
Furthermore, we demonstrate the importance of precise control of the
ac-Stark field intensity and unambiguously verify that intensity inhomogeneities
and resulting dephasing are the most important source of decoherence
in our system. With appropriate corrections we apply a
phase shift of over 15 rad within the
lifetime of the spin coherence. This shows
that a large variety of manipulations on spin-precession (traditionally denoted as FID) temporal
dynamics are possible, ranging from simple frequency shifts, through
inducing beat-notes, to rapid collapse-and-revival behavior. All of
this confirms that our platform is a versatile tool to efficiently
prepare complex spin patterns in the cold atomic ensemble.
\begin{figure}[!b]
\includegraphics[width=1\columnwidth]{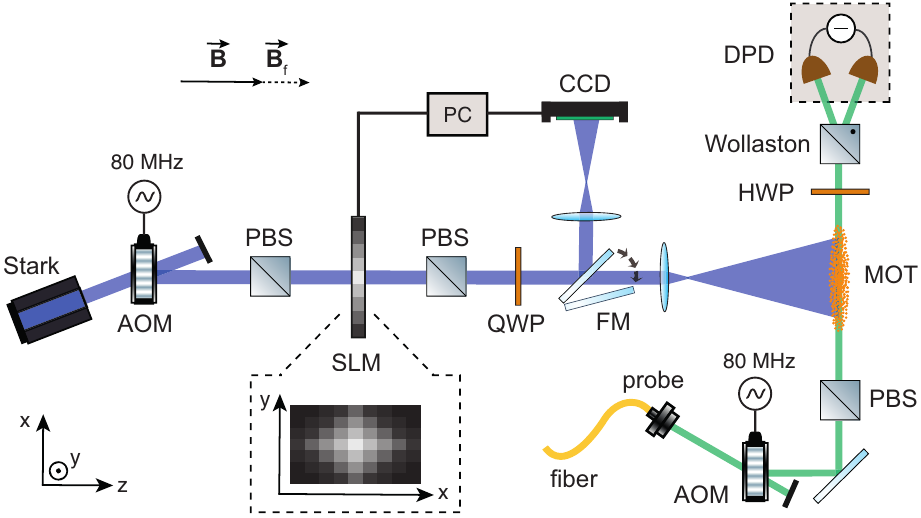}\centering
\caption{Schematic of the experimental setup. Atoms released from the magneto-optical
trap (MOT) are illuminated by two beams: linearly polarized probe
and circularly polarized ac-Stark beam. ac-Stark beam intensity at the
output of acousto-optic modulator (AOM) is shaped using reflective spatial light
modulator (SLM), drawn in the transmission configuration for simplicity. Flip mirror (FM)
switches SLM image between MOT and CCD camera. Differential photodiode
(DPD) registers polarization rotation of the probe light.  PBS, polarizing beamsplitter; HWP (QWP), half-wave (quarter-wave) plate.\label{fig:expScheme}}
\end{figure}

To accurately describe observed experimental results let us introduce
the basic theory of ac-Stark shift by considering the $F=1$ ground-state
manifold of $^{87}$Rb atom, for which the Hamiltonian describing interaction with $z$-propagating laser beam can be decomposed into three components \cite{Geremia2006,DeEchaniz2008,Colangelo2013}:
\begin{equation}
\hat{H}_{S}=\hat{H}_{S}^{(0)}+\hat{H}_{S}^{(1)}+\hat{H}_{S,}^{(2)}\label{eq:hamiltonian}
\end{equation}
with
\begin{equation}
\begin{array}{c}
\hat{H}_{S}^{(0)}=\frac{2}{3}g\alpha^{(0)}\hat{S}_{0},\\
\hat{H}_{S}^{(1)}=g\alpha^{(1)}\hat{S}_{z}\hat{F}_{z},\\
\hat{H}_{S}^{(2)}\! \! \! =\! g\alpha^{(2)}\! \! \left[\! \frac{1}{3}\hat{S}_{0}(3\hat{F}^2_{z}\! -\! 2\hat{\mathbb{1}})\! \! +\!\!  \hat{S}_{x}(\hat{F}_{x}^{2}\! -\! \hat{F}_{y}^{2})\! \! +\! \! \hat{S}_{y}(\hat{F}_{x}\hat{F}_{y}\! +\! \hat{F}_{y}\hat{F}_{x})\! \right]\! ,
\end{array}\label{eq:decompose}
\end{equation}
where: $\alpha^{(i)}$ is the tensor of atomic polarizability depending
on atomic quantum numbers and detuning $\Delta$, $g=\omega_{0}/(2\epsilon_{0}V)$ is a form factor \cite{Geremia2006} (with the resonant atomic frequency $\omega_{0}$ and the interaction
volume $V$), and $\hat{S}_{j}$ ($\hat{F}_{j}$) are Stokes (atomic spin) operators \cite{Geremia2006,DeEchaniz2008}.
For detuning $\Delta$ much bigger than the decay rate $\Gamma$,
tensor part of the interaction $\alpha^{(2)}$ is proportional to
$1/\Delta^{2}$, while scalar $\alpha^{(0)}$ and vector $\alpha^{(1)}$
parts are proportional to $1/\Delta$ \cite{DeEchaniz2008}. Furthermore, for strong circularly polarized light we have $\left|\langle\hat{S}_{z}\rangle\right|\gg\left|\langle\hat{S}_{x}\rangle\right|,\left|\langle\hat{S}_{y}\rangle\right|$.
Therefore the contribution of $\hat{H}_{S}^{(2)}$ can be neglected
for large detuning $\Delta$ and circular polarization. Moreover,
the scalar part $\hat{H}_{S}^{(0)}$ causes only constant shift of
energy levels which does not affect the atomic spin dynamics.

In consequence the only part
of the Hamiltonian $\hat{H}_{S}$ (Eq. (\ref{eq:hamiltonian})) we need to consider is the vector
term $\hat{H}_{S}^{(1)}$. Within the approximation of
a classical electromagnetism, the form factor $g$ obeys the relation:
$\hbar g\langle\hat{S}_{z}\rangle=qI/(2\epsilon_{0}c)$, where $I$
is the light intensity and $q=\pm1$ corresponds to $\sigma_{\pm}$ polarization
of the ac-Stark-inducing laser beam, so the final form of the vector Hamiltonian
is
\begin{equation}
\hat{H}_{S}^{(1)}=q\frac{\kappa}{\Delta}\frac{I}{2\hbar\epsilon_{0}c}\hat{F}_{z},
\end{equation}
where $\kappa=\alpha^{(1)}\Delta=\mathrm{const.}$ for $\Delta\gg\Gamma$.
This specific form is reminiscent of the Hamiltonian for an atom in
an external magnetic field applied along the $z$ direction, so we
can define a fictitious magnetic field \cite{Cohen-Tannoudji1972}:
\begin{equation}
\mathbf{B}_{\mathrm{f}}=q\frac{1}{g_{F}\mu_{B}}\frac{\kappa}{\Delta}\frac{I}{2\hbar\epsilon_{0}c}\hat{e}_{\mathbf{k}},\label{eq:bf-1}
\end{equation}
where: $\hat{e}_{\mathbf{k}}=\mathbf{k/|\mathbf{k}|}$ and $\mathbf{k}$
\textendash{} wavevector of the laser beam propagating along the $z$
direction. When atomic ensemble is also influenced by a real magnetic
field $\mathbf{B}$, taking into account all of the above approximations,
the total Hamiltonian $\hat{H}_{S}$ from Eq. (\ref{eq:hamiltonian}) can be written
in a traditional form:
\begin{equation}
\hat{H}=g_{F}\mu_{B}(\mathbf{B+B}_{\mathrm{f}})\mathbf{\mathbf{\hat{F}}.}\label{eq:bfield}
\end{equation}
In consequence, atomic spins exposed to the ac-Stark beam precess
around the effective magnetic field $\mathbf{B}_{\mathrm{eff}}=\mathbf{B}+\mathbf{B}_{\mathrm{f}}$
with the Larmor frequency $\omega_{\mathrm{L}}=g_{F}\mu_{B}B_{\mathrm{eff}}/\hbar$. 

The model presented here is derived neglecting incoherent excitation
of atoms and subsequent re-emission of light, including the $F=2$
manifold. In the leading order of perturbation calculation the rate
$\Gamma_{\mathrm{{scatt}}}$ of this incoherent scattering scales
as $\Gamma_{\mathrm{{scatt}}}\sim1/\Delta^{2}$ \cite{Sparkes2010}.
Therefore, in the far-detuned regime of our experiment ($\Delta\gg\Gamma$)
we expect this contribution to be insignificant. 
\begin{figure}[!b]
\includegraphics[width=1\columnwidth]{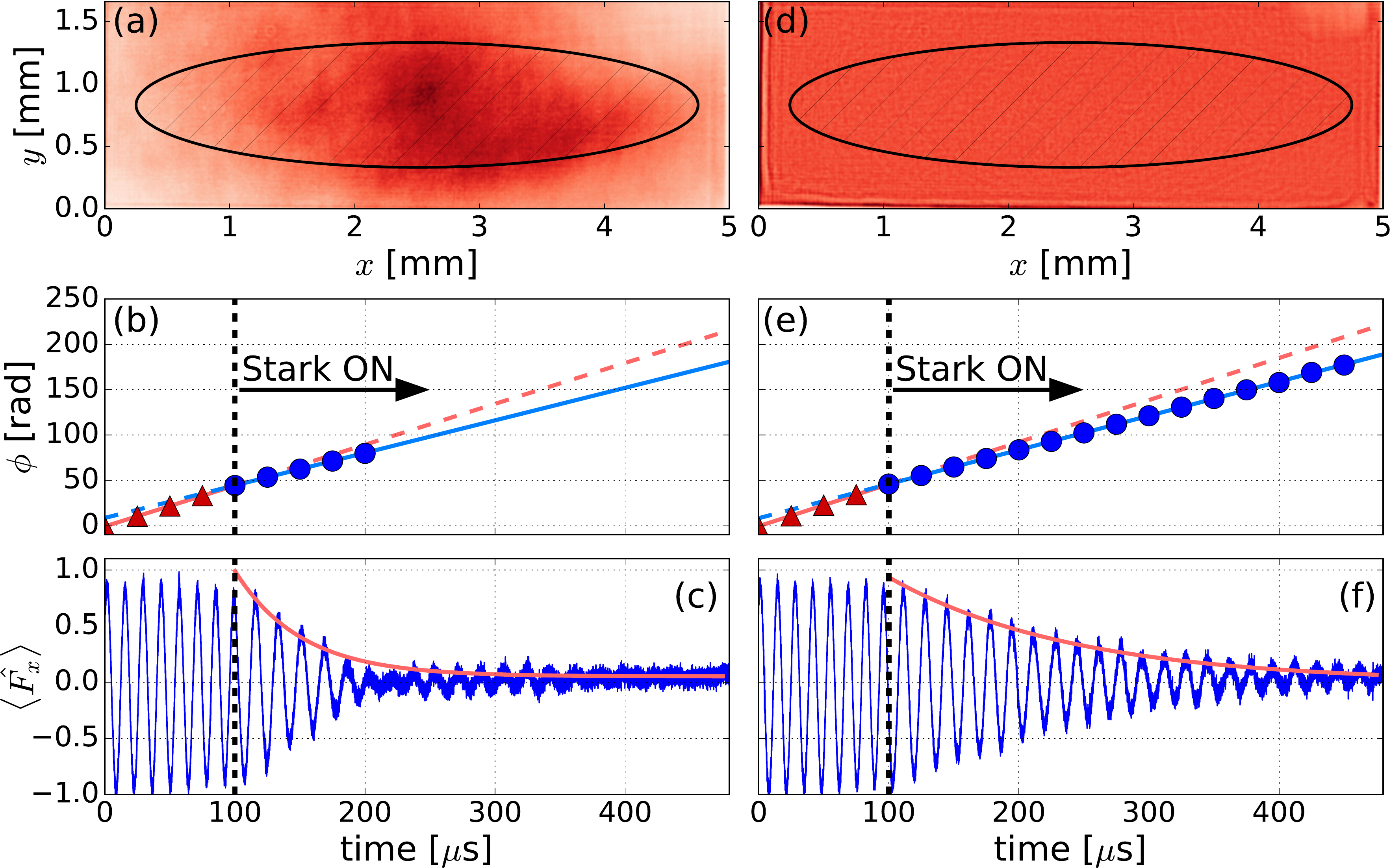}\centering
\caption{Influence of the ac-Stark beam intensity inhomogeneity on the FID lifetime $\tau$ for $B_{\mathrm{eff}}=B-B_{\mathrm{f}}$.
(a) Intensity distribution $I_{0}$ of $\sigma_{-}$-polarized ac-Stark
beam registered on the CCD camera, without SLM correction. The shaded ellipse visualizes MOT position. (b) Phase
$\phi$ retrieved from the FID signal oscillations presented (along with its envelope) in (c).
Triangles (dots) correspond to the phase $\phi$ measured before (after)
turning on the ac-Stark beam, along with the linear fits (errorbars within data points). Analogous
data for intensity distribution $I_{C}$, corrected using SLM to obtain
flat profile of the beam in the MOT plane, are presented
in (d)--(f). \label{fig:Influence-of-Stark}}
\end{figure}

In our experiment, presented in Fig. \ref{fig:expScheme}, we use
laser-cooled ensemble of $N=10^8$ Rb-87 atoms inside a magneto-optical trap (MOT)
(OD=40, for details see \cite{Parniak2017}). In a typical experimental sequence,
repeated synchronously with the 50 Hz SLM refresh rate, the atoms
are first cooled and trapped for 19.6 ms and then the MOT coils
are turned off to allow for 300 $\mu$s cooling in optical molasses,
so the ensemble can reach final temperature of 22 $\mu$K \cite{Parniak2017}. Residual magnetic fields from
eddy currents decay after 100 $\mu$s. Once the MOT is fully switched
off we pump the atoms to the $5^{2}S_{1/2},\:F=1$ state with $\langle F_{x}\rangle=1$.
Then, after 100 $\mu$s of atomic spins rotation driven only by an
external magnetic field $\mathbf{B}=B\hat{e}_{z}$ with amplitude $B=100$ mG, we illuminate
atomic ensemble with circularly polarized ac-Stark beam. The beam is
far-detuned from the $5^{2}S_{1/2},\:F=1\rightarrow5^{2}P_{3/2}$
atomic transition and propagates along the $z$ direction. This experimental
configuration results in the net magnetic field $\mathbf{B}_{\mathrm{eff}}=\mathbf{B}+\mathbf{B}_{\mathrm{f}}$
pointing along the $z$ axis.

Average spin projection $\langle\hat{F}_{x}\rangle$ onto the $x$ axis 
is measured by registering polarization rotation of a weak linearly
polarized probe beam propagating along the $x$ direction, using a
Wollaston prism and a differential photodiode (DPD) \cite{Geremia2006}. The probe beam of $1\:\mu\mathrm{W}$ power
is detuned by 100 MHz from the $5^{2}S_{1/2},\:F=1\rightarrow5^{2}P_{3/2},\:F=2$
transition, minimizing the deleterious effect of incoherent excitations
and tensor interaction. To avoid spin decoherence due to intensity
inhomogeneities of the ac-Stark beam we correct the spatial profile
using a phase-only reflective SLM and a polarizing
beamsplitter (PBS). The surface of the SLM is imaged on the atomic
ensemble (MOT) as well as onto a CCD camera 
situated in an auxiliary image plane of the SLM. This allows for
direct calibration of light intensity \cite{Leszczynski2015} incident on the atomic ensemble. With ray-trace modelling we estimate the spatial resolution of imaging the SLM surface onto the atomic ensemble to be $20\:\mu$m. The SLM imaging setup
is switched between CCD camera and MOT using a flip mirror (FM).

As the sign of the fictitious magnetic field $\mathbf{B}_{\mathrm{f}}$
changes with the light polarization (Eq. (\ref{eq:bf-1})), the value
of the effective magnetic field $B_{\mathrm{eff}}$ is the sum or
the difference of $B$ and $B_{\mathrm{f}}$. In Fig. \ref{fig:Influence-of-Stark}
the influence of the ac-Stark effect (with $\Delta=-2\pi\times30$ GHz) on the typical spin-precession
signal (FID) is presented for the case where
the fictitious magnetic field $\mathbf{B}_{\mathrm{f}}$ of amplitude $B_{\mathrm{f}}=20$ mG is subtracted
from the real magnetic field $\mathbf{B}$. Using standard Hilbert
transform method we retrieve the phase
and envelope of the measured FID signal. This allows to  recover only the essential parameters with high precision without fitting of the full sinusoidal FID
signal. In the left column of Fig. \ref{fig:Influence-of-Stark} we plot the spatial intensity profile
$I_{0}$ of the ac-Stark beam without any SLM correction procedure
(a), total accumulated phase of the FID signal (b), and the FID signal
itself (c). The right column (d)\textendash (f) portrays corresponding
data for the spatial intensity profile $I_{\mathrm{C}}$ already corrected
with the SLM. The average intensity of $160\:\mathrm{mW/cm^2}$ is selected so that the mean FID
frequency shift is the same for both $I_{\mathrm{C}}$ and $I_{0}$
intensity profiles. When the ac-Stark beam is applied, much shorter
lifetime $\tau$ is observed for the uncorrected, highly inhomogeneous case (Fig. \ref{fig:Influence-of-Stark}(c)). 
\begin{figure}[!b]
\includegraphics[width=1\columnwidth]{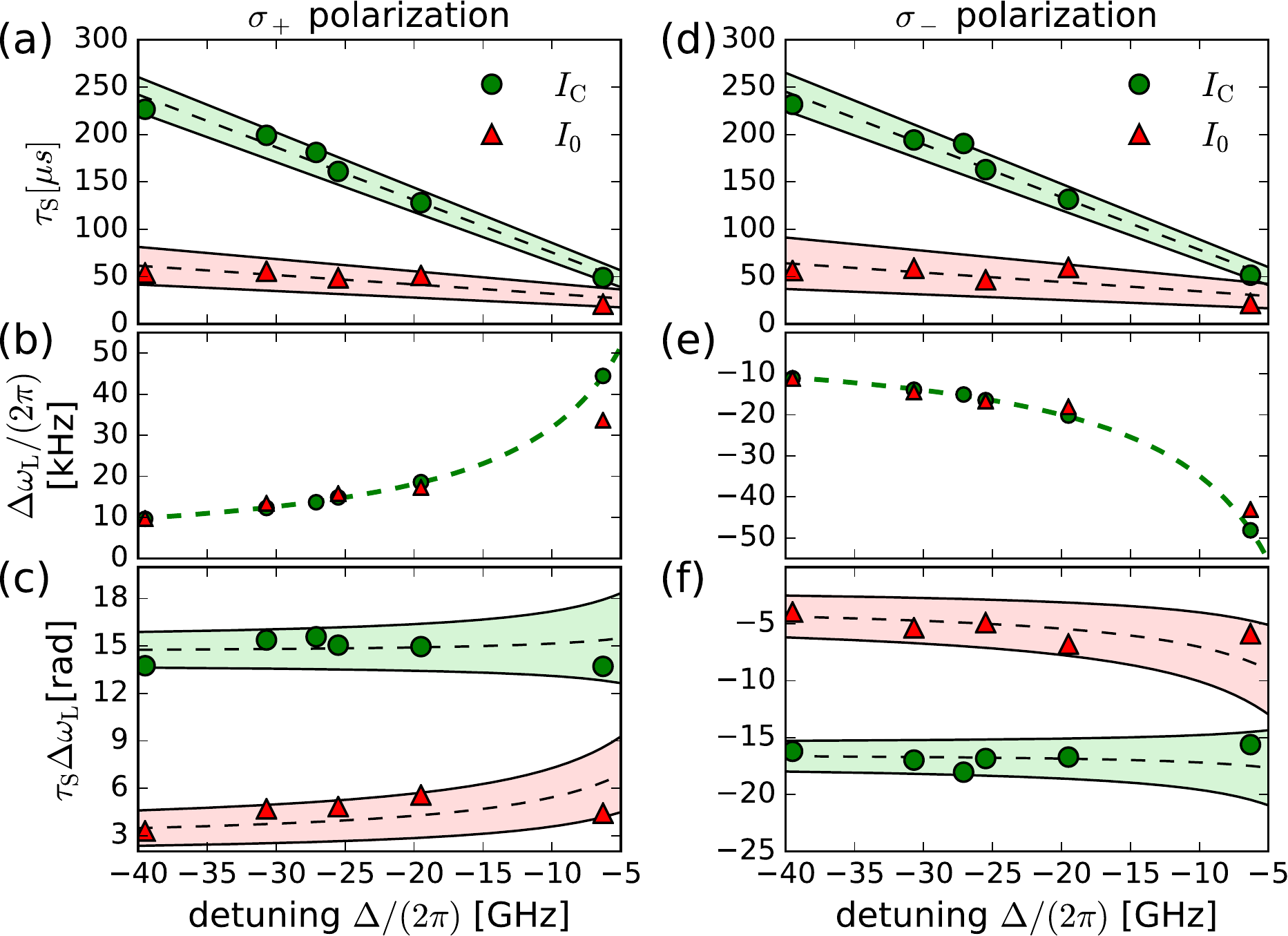}\centering
\textbf{\caption{Dependence of the FID signal on the detuning $\Delta$ for
the ac-Stark beam intensity profile with ($I_{C}$) and without ($I_{0}$)
SLM correction. Measured ac-Stark induced dephasing lifetime $\tau_{\mathrm{S}}$
along with a fitted function (dashed line) (a), Larmor frequency shift $\Delta\omega_{\mathrm{L}}$ (b) and total
phase $\phi_{\mathrm{S}}=\tau_{\mathrm{S}}\Delta\omega_{\mathrm{L}}$ (c)
accumulated within spin coherence lifetime (the product of values from (a) and (b)) for $\sigma_{+}$ polarization of the ac-Stark
beam. Shading regions around dashed lines (theoretical fit) correspond to fitting uncertainties (determined for (a) and (c) from standard covariance matrix of the linear fit parameters).
In (b) the theoretical scaling $I/\Delta$ is marked by the dashed line for intensity $I=160$ mW/cm$^{2}$ (errorbars within data points).
Analogous quantities for $\sigma_{-}$ polarization of the ac-Stark beam are plotted
in (d)\textendash (f). Maximum achievable frequency shift presented here corresponds to a fictitious magnetic field with amplitude $B_\mathrm{f}=70\ \mathrm{mG}$. \label{fig:Correlation-between-the}}}
\end{figure}

To explain the inherently finite lifetime $\tau$ of the spin-precession signal
let us now consider the three most essential sources of decoherence
for the case of ac-Stark shift spin control. The first type is any
kind of spin decoherence occurring even without the ac-Stark beam.
This includes dephasing due to magnetic field inhomogenities (occurs at a rate $< 1$ kHz thus negligible in further analysis) and most
importantly interaction of atoms with the probe beam, particularly
the tensor interaction (Eq. (\ref{eq:decompose})) which might be significant. Minimizing this
effect using weak probe light (ca. microwatt power) detuned by 100 MHz from the $5^{2}S_{1/2},\:F=1\rightarrow5^{2}P_{3/2},\:F=2$ transition, we obtain the FID lifetime (for $\omega_\mathrm{L}=2\pi\times74$ kHz) of about $\tau_{0}=700\ \mu$s. The second source \textendash{} associated with the manipulation
itself, is an absorption and subsequent re-emission of light caused by
the ac-Stark beam quantified by the scattering rate $\Gamma_{\mathrm{{scatt}}}$.
This effect is proportional to $I/\Delta^{2}$ \cite{Sparkes2010} while the Larmor frequency
behaves as $I/\Delta$ (Eq. (\ref{eq:bfield})), therefore the way to minimize incoherent scattering
is to increase $\Delta$ and $I$ proportionally. The last, significant
source of decoherence is the dephasing caused by the intensity inhomogeneities
of the ac-Stark laser beam characterized by the lifetime~$\tau_\mathrm{S}$, calculated as $1/\tau_\mathrm{S} = 1/\tau - 1/\tau_0$.

As shown in Fig. \ref{fig:Influence-of-Stark}, SLM-corrected ac-Stark
beam intensity distribution $I_{\mathrm{C}}$ increases the FID lifetime $\tau$
significantly compared to the situation with uncorrected intensity
profile $I_{0}$. To explicitly confirm this, we plot in Fig. \ref{fig:Correlation-between-the}
the lifetime $\tau_{\mathrm{S}}$ along with a fit to a simple relation $\tau_\mathrm{S} \propto |\Delta|$, ac-Stark induced Larmor frequency
shift $\Delta\omega_{\mathrm{L}}$ and phase accumulated within the FID lifetime
$\phi_{\mathrm{S}}=\tau_{\mathrm{S}}\Delta\omega_{\mathrm{L}}$, as
a functions of ac-Stark beam detuning $\Delta$. Two columns depict
results for both $\sigma_{+}$ and $\sigma_{-}$ polarized ac-Stark
light. The uncorrected
ac-Stark beam mean intensity is chosen to preserve the same FID frequency
as with corrected beam profile. Figure \ref{fig:Correlation-between-the}
(b) and (e) confirm the theoretical prediction $\Delta^{-1}$
 given by Eq. (\ref{eq:bf-1}). The curve fitted to
data corresponds to average light intensity $I=160$ mW/cm$^{2}$ which is
consistent with an independent measurement of light power. The total
phase $\phi_{\mathrm{S}}$ accumulated within FID lifetime exhibits
nearly $\Delta$-independent behavior which indeed confirms the dominant
role of the ac-Stark shift inhomogeneities on the decoherence phenomenon.
Most importantly, thanks to homogenization the possible phase $\phi_{\mathrm{S}}$ that
the spin can accumulate within one $1/e$ characteristic lifetime $\tau_\mathrm{S}$ increases from
$\phi_{\mathrm{S}}=5$ rad to about $\phi_{\mathrm{S}}=15$ rad for both
$\sigma_{+}$ and $\sigma_{-}$ polarizations of the ac-Stark beam.
It is more than enough to manipulate spin pattern in an arbitrary way. 
\begin{figure}[!b]
\includegraphics[width=1\columnwidth]{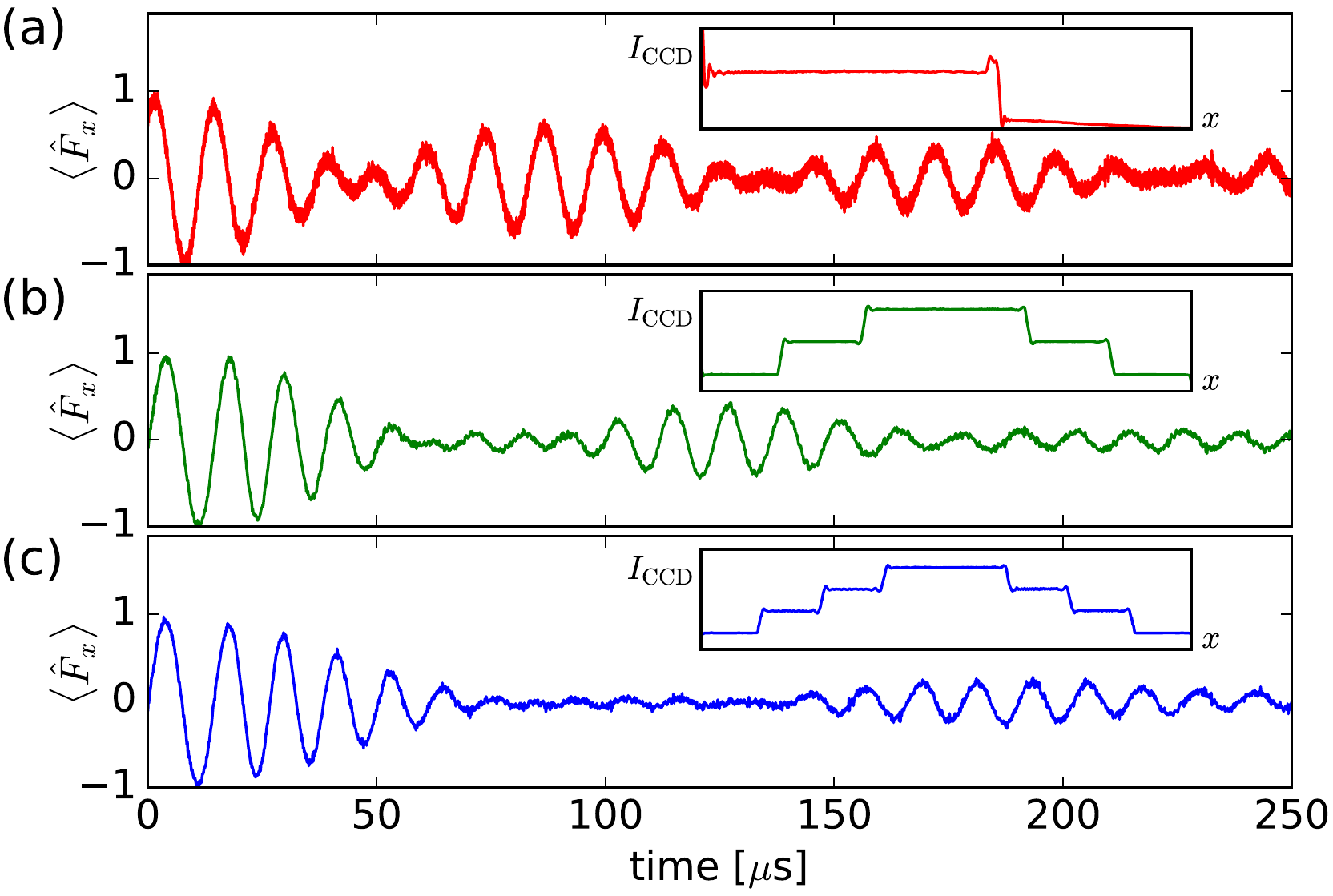}\centering
\caption{Temporal dynamics of the FID signal for staircase intensity profile $I_{\mathrm{CCD}}$
registered on the CCD camera.
Measured beat-note FID signal for two intensity steps (a) as well
as collapse-and-revival FID signal for three (b) and four (c) intensity
steps. Intensity patterns (insets) $I_{\mathrm{CCD}}$ are uniform in the $y$
direction on the CCD camera.\label{fig:FID-signal-for}}
\end{figure}

The manipulation of atomic spins using the ac-Stark effect (fictitious
magnetic field $\mathbf{B}_{\mathrm{f}}$) has many advantages over using only real magnetic field $\mathbf{B}$. First,
it gives us better temporal precision in the experiment, as using
the acousto-optic modulator (AOM) we can turn on/off the light in ca. hundred nanoseconds, which
is impossible in case of thr even low-inductance magnetic coils.
Second, using SLM we can easily sculpt the intensity $I_\mathrm{C}$ of the ac-Stark
beam into arbitrary shapes -- one particular example is depicted in
Fig. \ref{fig:FID-signal-for}. Here atoms are illuminated with the
staircase spatially-modulated ac-Stark beam (with $\Delta=-2\pi\times30$ GHz), visualized as cross-sections $I_{\mathrm{CCD}}$
of intensity profiles on the CCD camera. Thus,
several groups of spins oscillate with equidistant frequencies, forming
a frequency comb. With more intensity steps, or equivalently with
more teeth in the frequency comb, we can achieve lower ratio of the
revival duration to the time where the FID signal is collapsed, which scales linearly with the number of teeth. For only
two-intensity levels (Fig. \ref{fig:FID-signal-for} (a)) the FID
signal has a cosine envelope and collapses only for a moment, but for
four-level staircase (Fig. \ref{fig:FID-signal-for} (c)) the FID
signal almost completely disappears for about 80 $\mu$s. 

To conclude, we have demonstrated spatially-resolved control of the fictitious
magnetic field generated using the ac-Stark effect. We have shown that inhomogeneities
of the ac-Stark beam are the main source of dephasing. After homogenization
using SLM we have achieved a phase shift of over 15 rad within the $1/e$
lifetime of the spin coherence, which is several times longer than
without spatial intensity corrections. We have also presented the possibility to engineer complex temporal
dynamics of the FID signal by sculpting the ac-Stark beam intensity, from
frequency shift through beat-note, to collapse-and-revival behavior.
The control over spatial and temporal aspects of light in comparison
to real magnetic field makes our method very robust and useful in
high-resolution magnetometry \cite{Koschorreck2011}, magnetic gradiometry
\cite{Deb2013} or spin wave manipulations \cite{Sparkes2010}, giving prospects to readily improve the sensitivity and precision.

\paragraph*{Funding.}
Narodowe Centrum Nauki (NCN) (2015/19/N/ST2/01671, 2016/21/B/ST2/02559, 2017/25/N/ST2/01163 and 2017/25/N/ST2/00713); Ministerstwo Nauki i Szkolnictwa Wy\.{z}szego (MNiSW) (DI2013 011943 and DI2016 014846).

\paragraph*{Acknowledgment.}
The authors thank M. Jachura and M. Semczuk for careful reading the manuscript as well as K. Banaszek for generous
support.

% Bibliography

\end{document}